\documentstyle[12pt,epsf,epsfig]{article}
\newcommand {\ebar}{\hbox{E\kern-0.5em\lower-0.1ex\hbox{/}}}

\begin{document}

\begin{titlepage}

~~~~~~~~~~~~~~~~~~~~~~~~~~~~~~

\vspace{2cm}
\begin{flushright}
LBNL-39176
\end{flushright}

\begin{center}
{\Large\bf Fragmentation Functions approach in pQCD fragmentation phenomena}\\
\vspace{2cm}
{\large Simona Rolli}\\
\vspace{0.7cm}
{\it INFN-Pavia, Italy and Lawrence Berkeley Laboratory, Berkeley, USA}\\
{\it E-mail: rolli@fnal.gov}


\end{center}

\vspace{3cm}

\begin{center}

\begin{abstract}
Next-to-leading order parton fragmentation functions into light mesons are presented. 
They have been 
extracted from real and simulated $e^+e^-$ data and used to predict inclusive
single particle distributions at different machines.
\end{abstract}

\end{center}

\vspace{6cm}

\begin{center}
Talk given at the XXXI Rencontres de Moriond, QCD and hadronic
interactions,\\ 
Les Arcs, France March 24-31 1996
\end{center}

\end{titlepage}

\section{Introduction}

One of the fundamental properties of QCD is the shrinking of the coupling constant as the 
energy of the interactions grows (asymptotic freedom). This implies that perturbative 
techniques can be used to study high energy hadronic or leptonic phenomena.
In spite of this, we can not rely completely on perturbation theory because the fundamental
particles whose interactions become weak at high energy are deeply bound inside the hadrons
we use as beams, targets or as observables. The solution is given by the factorization
theorems, whereby cross sections can be expressed as the products of factors, each one
involving phenomena appearing at different energy scales.
A generic hadronic collisions with production of one hadron inclusive can be expressed
as the convolution of the partonic hard cross section and structure and fragmentation
functions, which represent respectively the parton density inside the hadrons and the
hadron density inside the parton.
Those parton model distributions acquire a scale dependence when the QCD corrections are 
included, the scale being the point where collinear initial (final) state divergencies
are subtracted.  Moreover these distribution are indipendent of the specific reaction.
The universality of these distributions is a key property, since they are not calculated 
from first principles as they contain non-perturbative information. Nevertheless they
evolve with the scale following Alterelli-Parisi type equations and this permits to  
extract them from a low-energy process and use them to predict rates for another one.

In this talk I'll concentrate on some recent sets of parton fragmentation functions into
light mesons. These sets have been extracted from real and simulated $e^+e^-$ data in a 
full NLO formalism. They have been used to predict inclusive single particle differential
distributions
at different energies and machine and to study jet fragmentation properties.
The outline of the talk is the following: I'll describe briefly how the sets have been
derived, then I'll give some results on the inclusive single particle predictions
comparing with data when they are available.

\section{Fragmentation functions from $e^+e^-$ data}

Parton fragmentation functions are defined as the density probability to find a hadron
$h$ inside a parton $p$. They are parametrized in the following form:
\begin{equation}
D_p^h (x,M_f) = N x^{\alpha} (1-x)^{\beta}
\end{equation}
where $x$ is the fraction of parton momentum carried by the hadron, while $M_f$ is the 
scale the function is evaluated. They satisfied multiplicity, charge and momentum 
conservation sum rules and evolve following Altarelli Parisi type equation. This means 
that given a certain 
set of input functions  at the scale $M_{f0}^2$ 
we can obtain the functions at any desired scale simply solving the AP equation.
To obtain the input sets we need to rely on experimental informations, because FF are
not calculable from first principle, owing the fact that they contain non-perturbative
informations\footnote{ we can actually calculate in a completely indipendent way only
heavy quarks fragmentation functions, because in this case the mass of the heavy
object act as a natural cutoff.}. 

To extract the FF we use $e^+e^-$ data. In the language of the QCD-improved parton 
model, in fact,  the $x$ distribution of the process $e^+e^-\to h+X$ emerges from the $x$ 
distribution $(d\sigma/dx)(x,\mu^2,s)$ of $e^+e^-\to a+X$ through convolution with
$D_a^h(x,M_f^2)$:

{\scriptsize
\begin{equation}
{1\over {\sigma_{TOT}}}{d\sigma(e^+e^-\to h+X)\over{dx}}=\sum_a{}\int_x^1{dz\over z}
D^h_a(z,M_f^2){1\over {\sigma_TOT}}{d\sigma_a\over{dy}}\left({x\over z},\mu^2,M_f^2\right);
\end{equation}
}
where $\sigma_{TOT}=N_c\sum_{i=1}^{N_f}{}e_q^2\sigma_0$ and

{\scriptsize
\begin{equation}
{1\over {\sigma_{TOT}}}{d\sigma_{q_i}\over{dy}}\left(y,\mu^2,M_f^2\right)=
e_q^2N_c{\sigma_0\over{\sigma_{TOT}}}
\left\{\delta(1-y) + \frac{\alpha_s(\mu^2)}{2 \pi} \left[
P^0_{qq}(y) \ln\!\left( \frac{s}{M^2_f} \right) + K_q^T(y) 
+K^L_q(y) \right]\right\}
\end{equation}
}
and

{\scriptsize
\begin{equation}
{1\over {\sigma_{TOT}}}{d\sigma_{g}\over{dy}}\left(y,\mu^2,M_f^2\right)=
\frac{2\alpha_s(\mu^2)}{2 \pi} \left[
P^{(0,T)}_{qg}(y) \ln\!\left( \frac{s}{M^2_f} \right) + K_g^T(y) 
+K^L_g(y) \right]
\end{equation}
}

\begin{figure}[htb]
\begin{center}
\begin{minipage}{7cm}
\epsfxsize=5cm
\epsffile{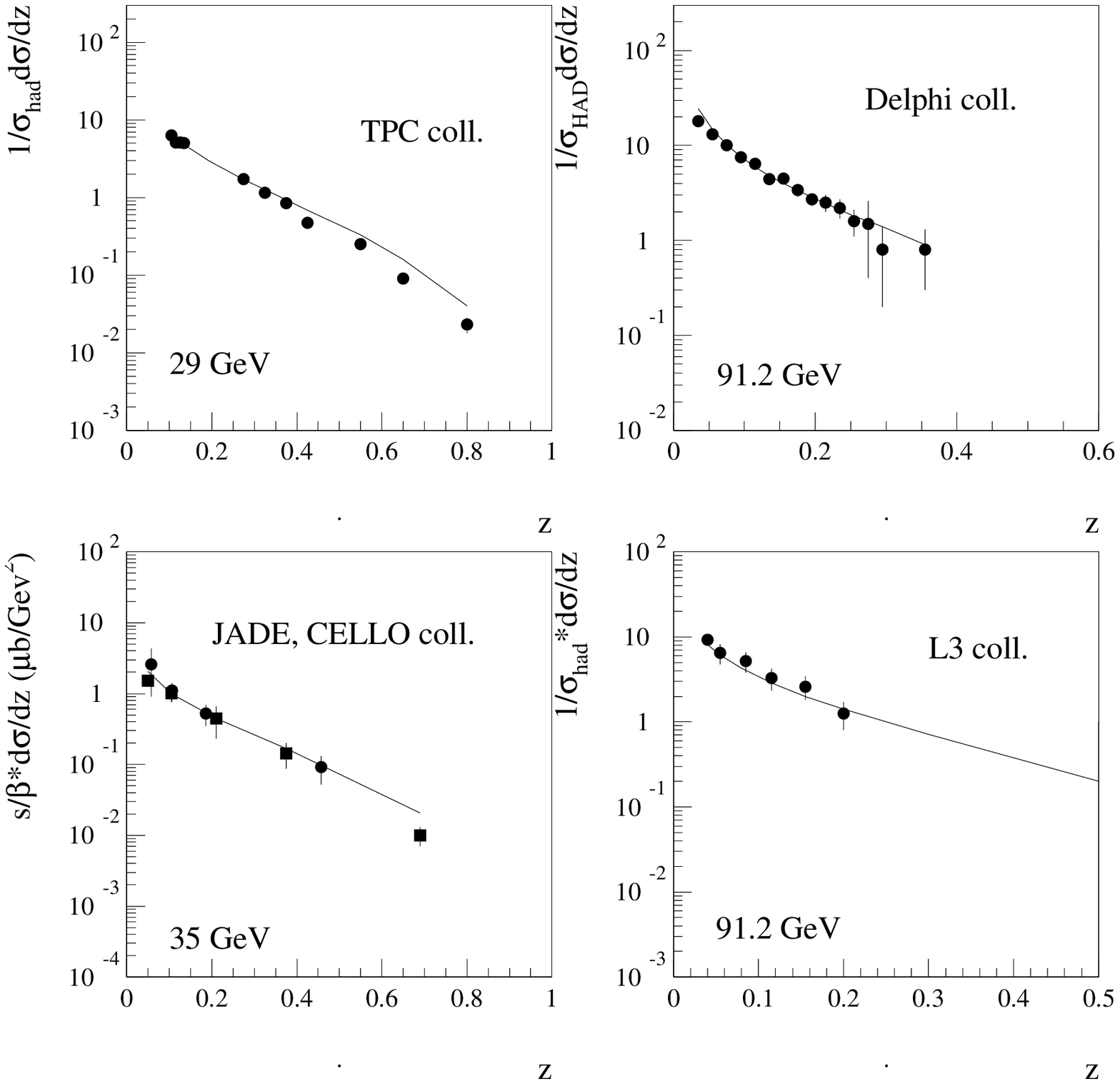}
\caption{{\scriptsize $K^{\pm}$ (upper right), $K_s^0$ (upper left) and $\eta$ (lower)
production using sets from ref.[6,8]}  }
\label{fig1}
\end{minipage}
\hfill
\begin{minipage}{7cm}
\epsfxsize=5cm
\epsffile{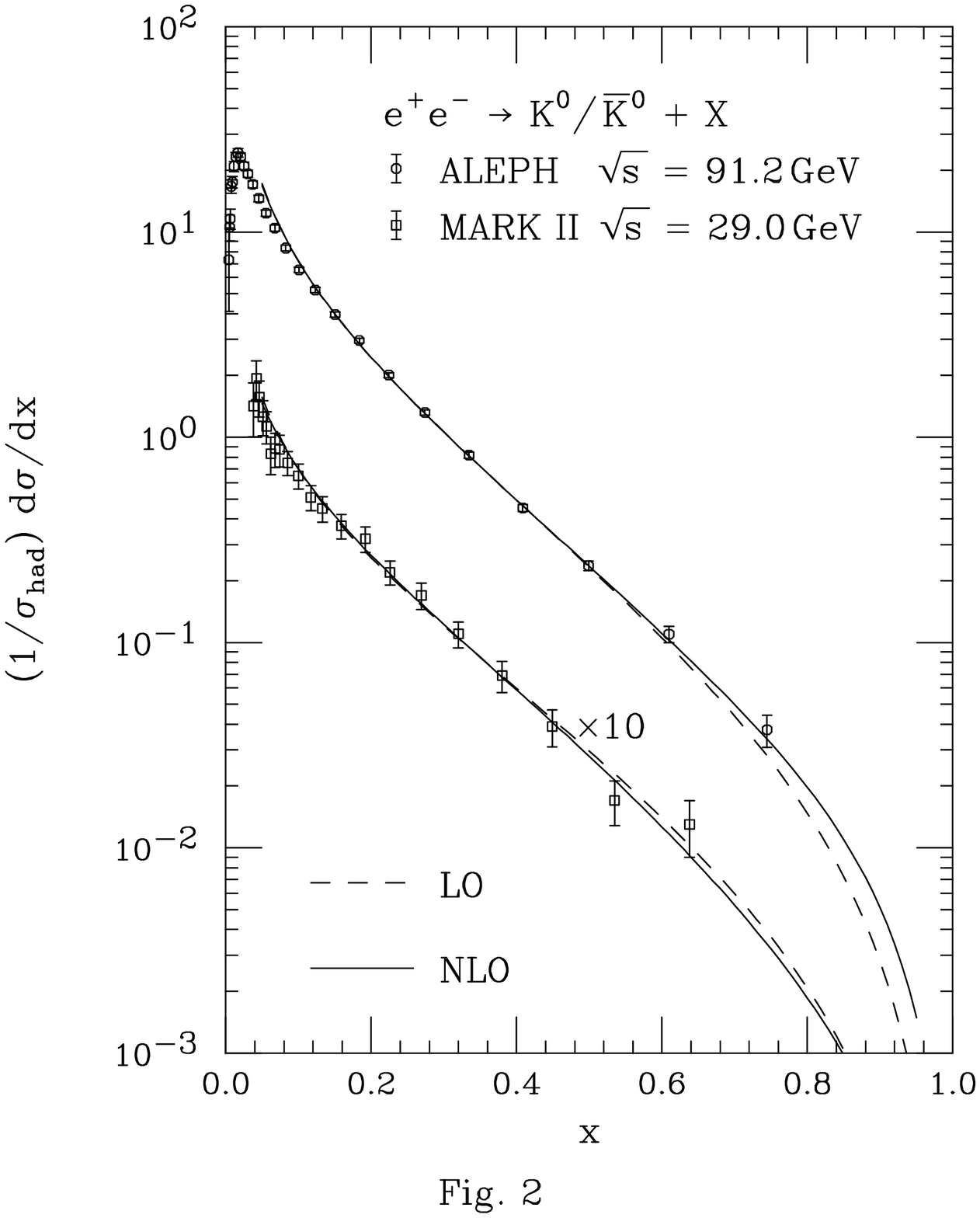}
\caption{{\scriptsize neutral kaons production at $e^+e^-$ using the sets from 
ref.[5] }}
\label{fig2}
\end{minipage}
\end{center}
\end{figure}

The functions $K^T_q$, $K^L_q$, $K^T_g$ and $K^L_g$ are shown
in reference \cite{simo1} and represent the NLO corrections.
Using real and simulated\cite{webber} data several different sets have been derived for 
neutral\cite{simo1} and charged pions\cite{simo2,kramer1}, 
neutral and charged kaons\cite{simo3,kramer2} and $\eta$ mesons\cite{simo4} at different
reference scales. In Figure 1 and 2 we give some examples
on how well the different sets, when convoluted with the
partonic cross sections, agree with $e^+e^-$ data. We refer the readers to the previous
references for further details.

\section{Applications}

As already stated, fragmentation functions are process independent:
this means that we can use them to make predictions for process different from
the one we used to extract them. We
simply need to convolute them with the appropriate partonic cross sections.
In the previous references the different sets have been used
to give predictions on inclusive single particle production in 
hadron-hadron, hadron-lepton and photon-photon\cite {kramer4} collisions, 
and to phenomenologically describe jet fragmentation properties.
In Figures 3, 4 and 5 we show again some examples on
the NLO predictions for different energy ranges and process types.

\begin{figure}[htb]
\begin{center}
\begin{minipage}{6.5cm}
\epsfxsize=5.5cm
\epsffile{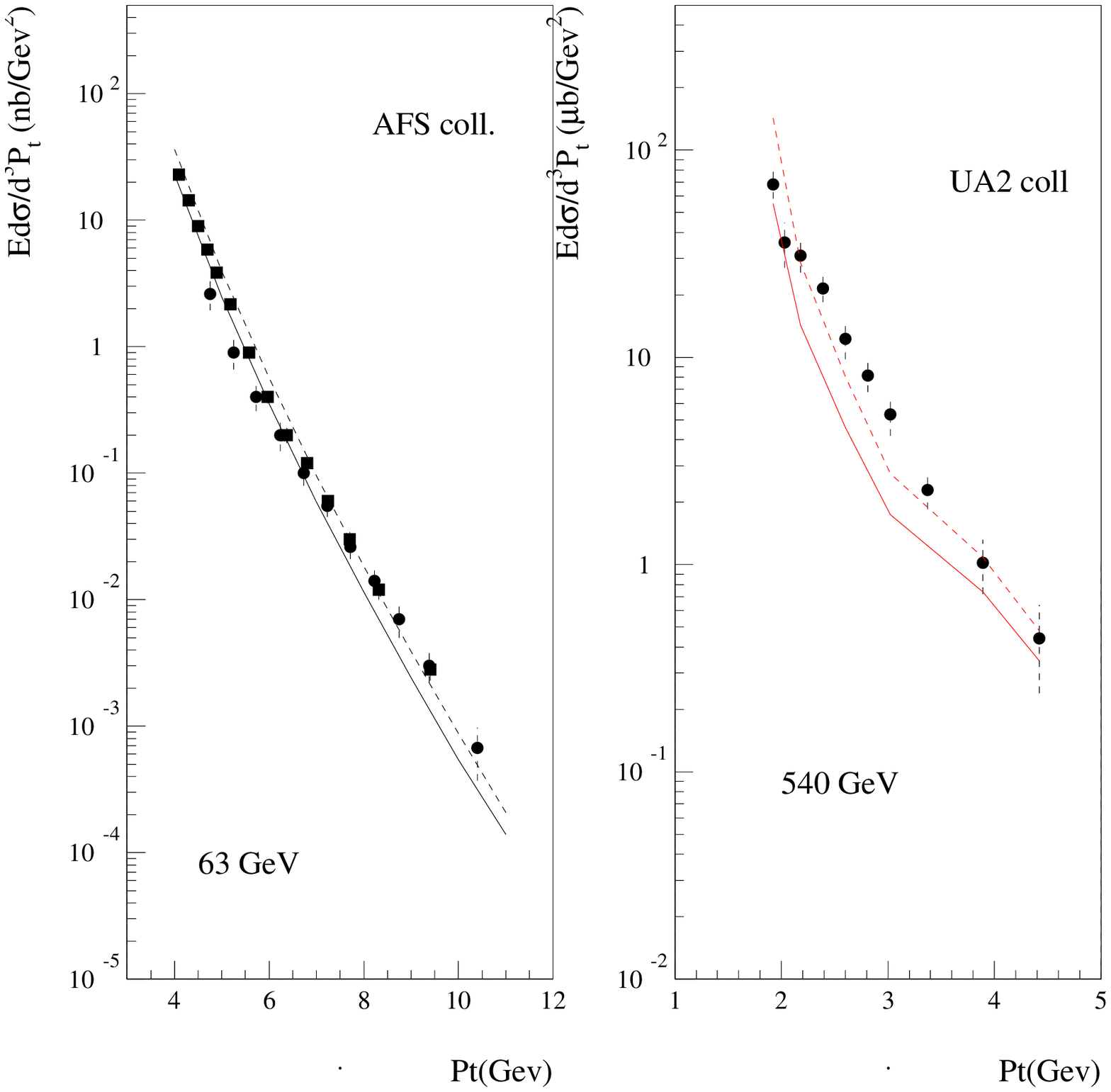}
\caption{{\scriptsize $\pi^0$ production at ISR and SppS using sets from
ref.[1]}}
\label{fig3}
\end{minipage}
\hfill
\begin{minipage}{6.5cm}
\epsfxsize=5.5cm
\epsffile{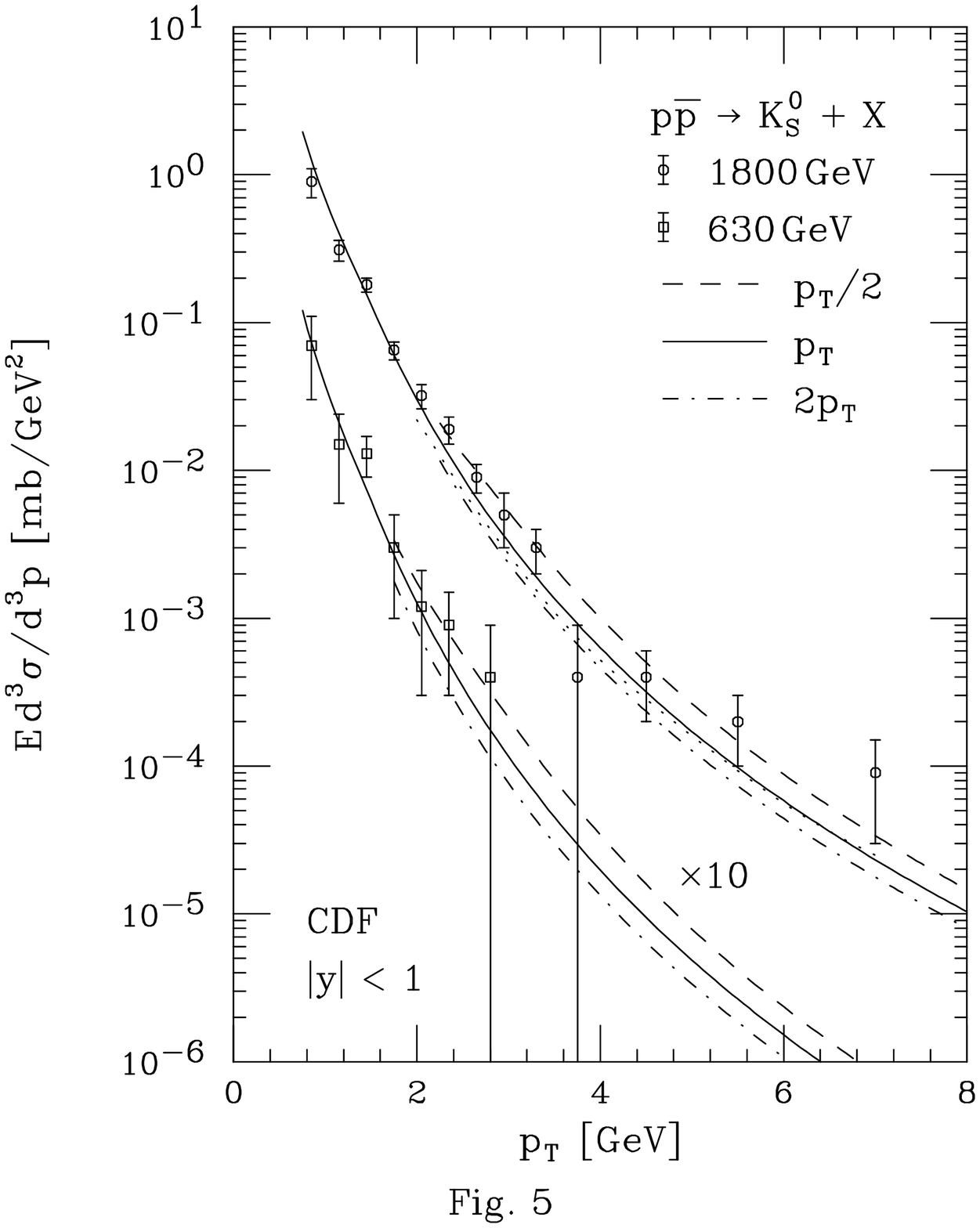}
\caption{{\scriptsize $K_s^0$ production at TeVatron using
sets form ref.[5]. The dotted line is from ref.[6] }}
\label{fig4}
\end{minipage}
\end{center}
\end{figure}

Finally,
in order to disentangle the fragmentation properties and the hadronization 
mechanism of high $p_t$ jets, we consider the ratio between 
the single hadron and jet cross sections, for fixed values of the variable
$z=E_{hadr}/E_{jet}$. Then, using the cone  jet algorithm 
and the NLO evaluation of the jet cross sections of
ref. \cite{aversa},  we present   
in Figure 6 a result on jet fragmentation in 
charged and neutral pions, with the energy of the jet varying between 40 and 
70 GeV, and a jet cone radius R=0.7 centered around the $\eta=0$ direction.
The overall theoretical uncertainty -which is not reported in figure- 
can be estimated to be of order 50\%. 
We also show the analogous
experimental result on jet fragmentation in charged hadrons from CDF, 
in reasonable
agreement with the theoretical prediction.

\begin{figure}[htb]
\begin{center}
\begin{minipage}{6.cm}
\epsfxsize=4.cm
\epsffile{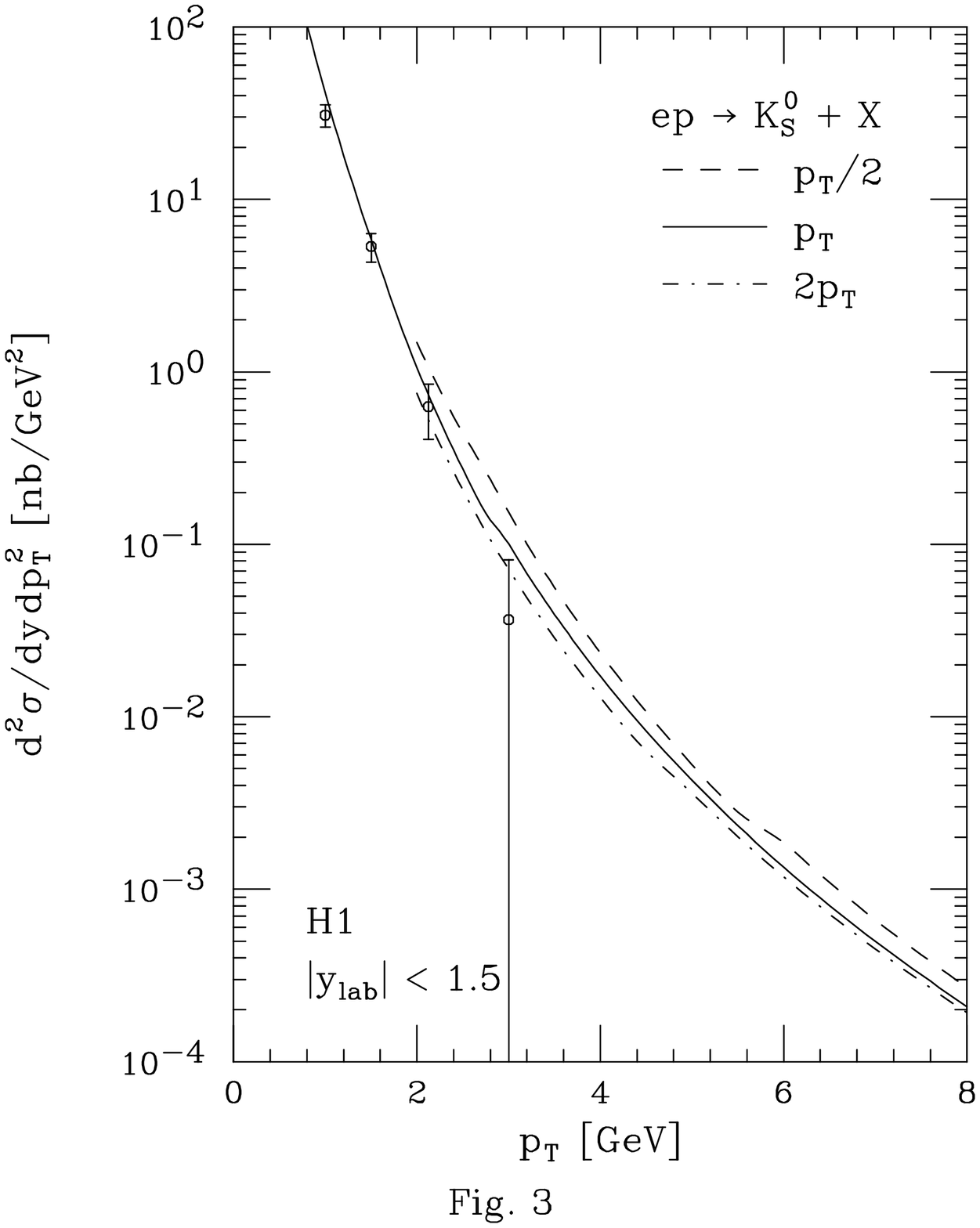}
\caption{ {\scriptsize $K_s^0$ production at HERA,
 using sets from ref.[5] }} 
\label{fig5}
\end{minipage}
\hfill
\begin{minipage}{6cm}
\epsfxsize=4.cm
\epsffile{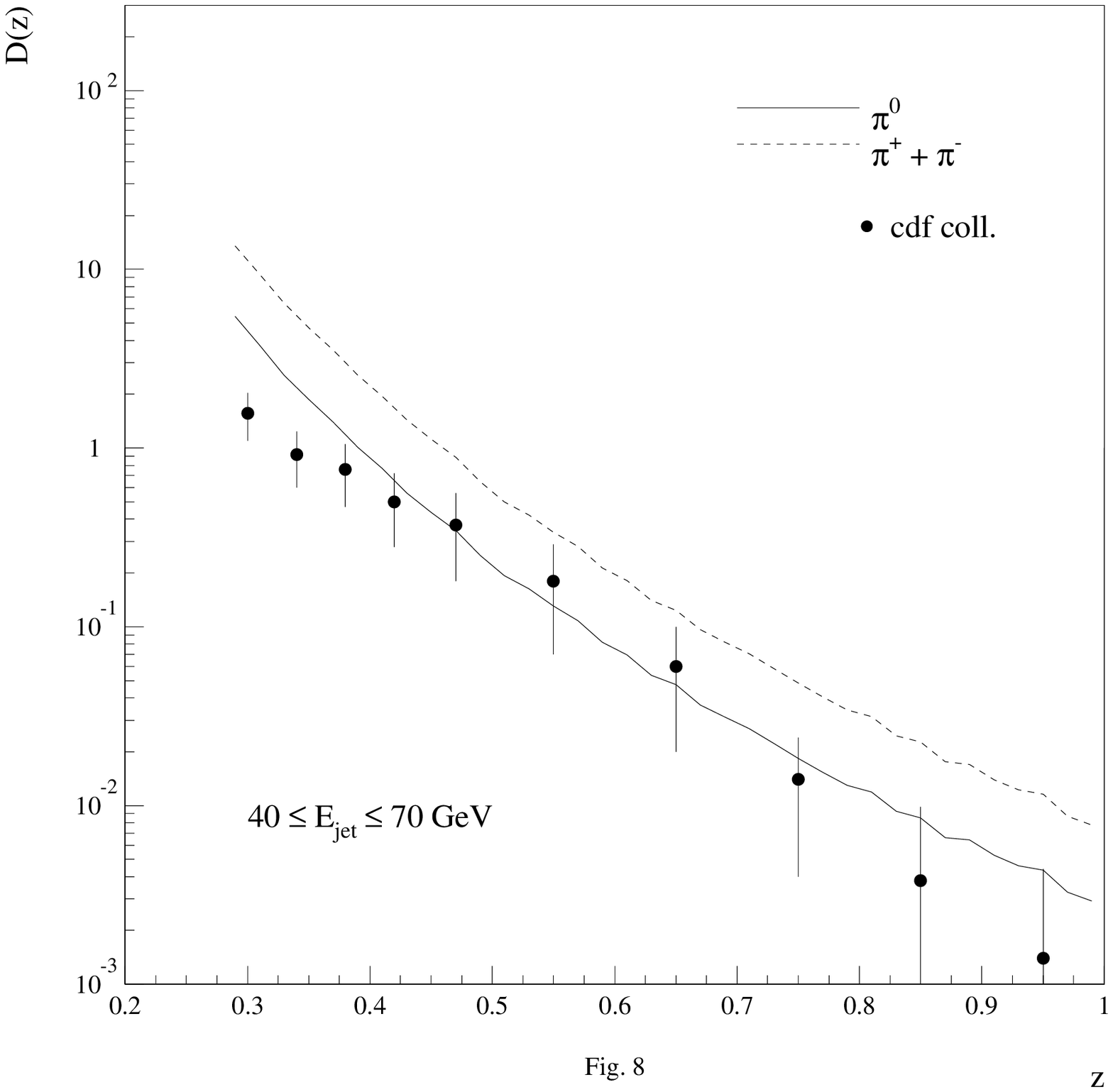}
\caption{{\scriptsize jet fragmentation function, into $\pi^0$
 and $(\pi^+ + \pi^-)$. The experimental points refer 
to charged hadrons  and are from CDF }}
\label{fig6}
\end{minipage}
\end{center}
\end{figure}

\vfill\eject

\section{Conclusions}

In this talk I reviewed the results on some recent NLO sets of parton
fragmentation functions into light mesons. The possibilities to use them to make
reliable predictions at different type of machines has been briefly shown.

\section{Acknowledgement}
This work was supported in part by the ``Maria Rossi'' fellowship from Collegio Ghislieri of Pavia,
through the U.S. Department of Energy under contract DE-AC03-76SF00098.

\end{document}